\documentclass[aps,prl,twocolumn,floatfix]{revtex4}
\usepackage{bm}
\usepackage{epsf}
\usepackage{color}
\usepackage{amssymb}
\usepackage{amsmath}
\usepackage{graphicx}
\usepackage{rotating}
\usepackage{epsfig}
\usepackage{psfrag}
\usepackage{amsmath}
\usepackage{hyperref}
\usepackage{subfigure}
\newcommand{\bk}{{\bf k}}

\newcommand{\gtappr}{{{\lower4pt\hbox{$>$} } \atop \widetilde{ \ \ \ }}}

\begin{document}

\title{Kondo Breakdown and  Quantum Oscillations in SmB$_6$}

\author{Onur Erten$^1$,  Pouyan Ghaemi$^2$ and Piers Coleman$^{1,3}$ }
\affiliation{$^1$Center for Materials Theory, Rutgers University, Piscataway, New Jersey, 08854, USA \\ 
$^2$ Physics Department, City College of the City University of New York, New York, NY 10031 \\$^3$Department of Physics, Royal Holloway, University of London, Egham, Surrey TW20 0EX, UK}

\begin{abstract}
Recent quantum oscillation experiments on SmB$_6$ pose a paradox, for
while the angular dependence of the oscillation frequencies suggest
a 3D bulk Fermi surface, SmB$_6$ remains
robustly insulating to very high magnetic fields. Moreover, a sudden low
temperature upturn in the amplitude of the oscillations 
raises the possibility of quantum criticality.  Here we
discuss recently proposed mechanisms for this effect, contrasting bulk and surface scenarios.  
We argue that topological surface states
permit us to reconcile the various data with bulk
transport and spectroscopy measurements,  interpreting the low
temperature upturn in the quantum oscillation amplitudes as a result
of surface Kondo breakdown and the high frequency
oscillations as large topologically protected orbits 
around the X point. We discuss various predictions that can be 
used to test this theory.

\end{abstract}
\maketitle
SmB$_6$, discovered 50 years ago\cite{Vainshtein:1965wh,Menth_PRL1968}, 
has  attracted recent 
interest due to its unusual surface transport properties:
while its insulating gap develops around $T_K\simeq
50$K, the resistivity saturates below a few Kelvin\cite{Allen_PRB1979}.
The renewed interest derives in part from 
from the possibility that SmB$_{6}$ 
is a topological Kondo insulator, developing topologically protected 
surface states at low temperatures
\cite{Dzero_PRL2010, Dzero_PRB2012,Alexandrov_PRL2013, Fu_PRL2013}. 
Experiments\cite{Wolgast_PRB2013,
 Kim_SciRep2013, Kim_NatMat2014} have confirmed 
that the plateau conductivity derives from surface states, and these 
states  have been resolved by angle-resolved photoemission spectroscopy
(ARPES)\cite{Jiang_NatComm2013,Neupane_NatComm2013, 
Xu_PRB2013, Frantzeskakis_PRB2013}.  Furthermore, spin-ARPES experiments 
have revealed the spin-momentum locking of the surface quasiparticles
expected from topologically protected Dirac cones\cite{Xi_NatComm2014}. 

Yet despite this progress, some important
experimental results are unresolved.  In particular, 
quantum oscillation experiments on SmB$_6$ have given rise to two
dramatically different interpretations\cite{Li_Science2014, Sebastian_Science2015}. Ref
\cite{Li_Science2014} observes low frequency (small Fermi
surface) oscillations 
with the characteristic $1/\cos(\phi)$ dependence on field orientation
expected from 2D topological surface states. On the other hand Ref
\cite{Sebastian_Science2015} detects a wide range of frequencies (both
high and low frequency oscillations) which have been interpreted in
terms of angularly isotropic three dimensional quasiparticle orbits,
resembling a metallic hexaboride without a
hybridization gap (such as LaB$_6$). A striking aspect of these
measurements, is that the 
oscillations strongly deviate from a classic Lifshitz-Kosevich formula below
$\sim1$K.  Two
recent
theoretical proposals have been advanced to
account for this bulk behavior, as a consequence of magnetic
breakdown\cite{Knolle_Arxiv2015}, of the formation of non-conducting
Fermi surfaces\cite{Baskaran_Arxiv2015}.
\begin{figure}[t!]
\centerline{
\includegraphics[width=7cm]{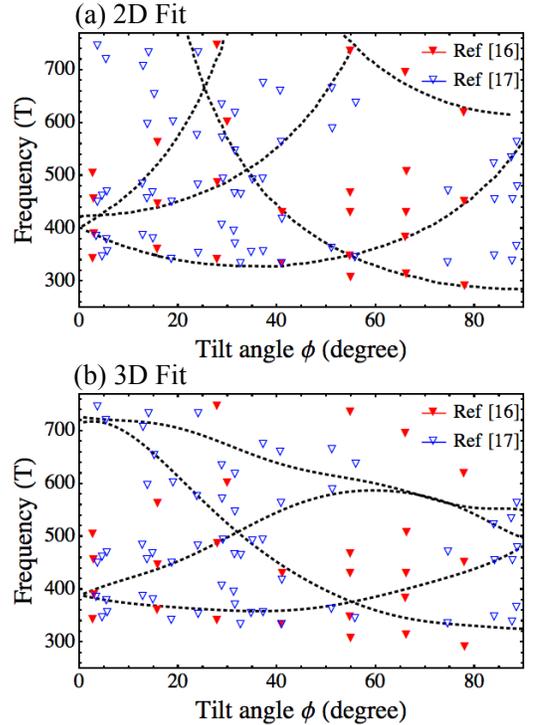}
}
\caption{Low frequency oscillations extracted and replotted from Ref\cite{Li_Science2014} (red, filled symbols) and Ref\cite{Sebastian_Science2015} (blue, open symbols). Lines are model fits based on (a) two dimensional surface states\cite{Li_Science2014}, (b) three dimensional bulk bands\cite{Sebastian_Science2015}. Error bars are the size of symbols which include both experimental errorbars and also errors from extracting data from a logarithmic plot.
Given the error bars, there is a good agreement between two experiments for low frequencies and it is not possible to distinguish if a 2D or a 3D model fits better {\sl \' a priori}.}
\label{Fig:1}
\end{figure}

\begin{figure*}[t!]
\centerline{
\includegraphics[width=17cm]{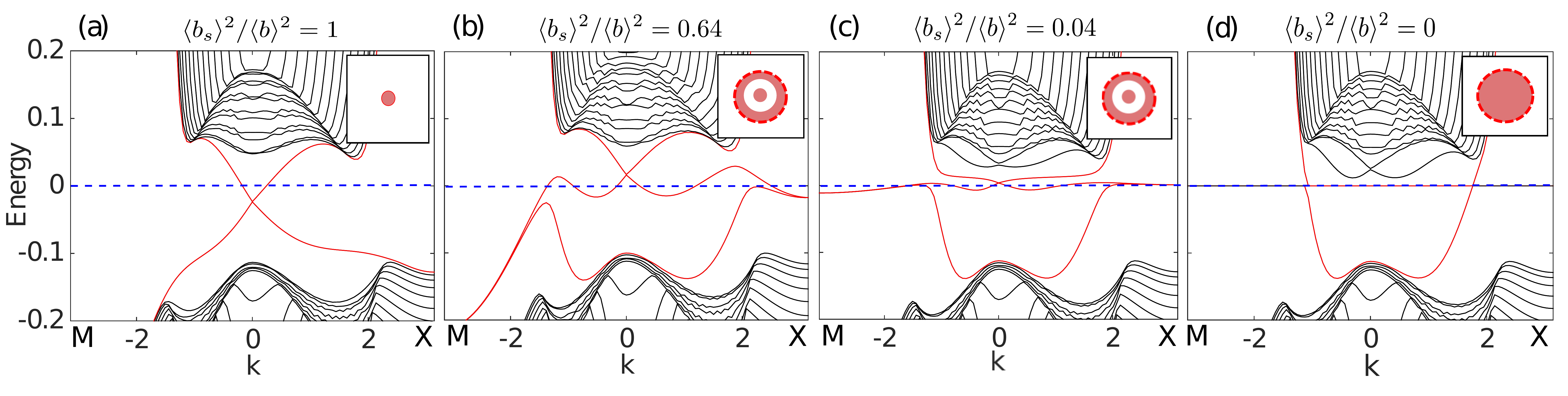}
}
\caption{Kondo breakdown and onset of surface Kondo effect. 3D strip calculations are done for varying $\langle b_{(s)}\rangle^2/\langle b\rangle^2$ where $\langle b_{(s)}\rangle^2$ is the bulk (surface) slave-boson amplitude indicating the amount of hybridization between the local moments on the surface and the topological surface states. For (a), $\langle b_s\rangle^2 /  \langle b\rangle^2 =1$, the hybridization on the surface is the same as the bulk, leading to heavy surface quasiparticles with a small Fermi surface. As $\langle b_s\rangle^2 / \langle b\rangle^2$ gets smaller in (b) and (c), the Fermi surface area increases as a result of Luttinger's sum rule and a new larger Fermi surface orbit appears. For  $\langle b_s\rangle^2 /  \langle b\rangle^2 = 0$ (d), local moments decouple from the surface, giving rise to a large Fermi surface with light quasiparticles. Insets show the Fermi surfaces for each figure.} 
\label{Fig:2}
\end{figure*}

Another aspect of recent measurements, is the wide disparity in
the reported effective masses of the carriers. 
The effective mass observed in both quantum
oscillation experiments, $m^*/m \sim 0.1-0.2$ is an order of magnitude
smaller than effective mass observed in
ARPES\cite{Jiang_NatComm2013,Neupane_NatComm2013, Xu_PRB2013,
Frantzeskakis_PRB2013}, $m^*/m \sim 1 $ and two orders of magnitude
smaller than those extracted from magnetothermoelectric
transport experiments,\cite{Luo_PRB2015} $m^*/m \sim 10-100$. Moreover
the Fermi surface areas are also drastically different in these sets of
different experiments.

The goal of this paper, is to discuss and contrast the quantum
oscillation data that has inspired these alternative interpretations,
seeking an interpretation that can be reconciled with the robust
insulating behavior and surface conductivity of this Kondo insulator.
As shown in Fig. 1, in the low frequency range where the two sets of
experiments overlap, the data are in agreement within the error bars,
and {\sl a priori} permit either a two, or three dimensional
interpretation.  Beginning with an examination of possible three
dimensional bulk interpretations of the data, we will argue that
consistency obliges us to return to a surface interpretation of the
quantum oscillations, derived from 2D topological surface states with
a $1/\cos(\phi-\phi_{0})$ angular dependence off various surface
facets that can resemble the isotropy of three dimensional orbits.  In
our theory, the large orbit quantum oscillations observed in
experiment are a consequence of the suppression of the Kondo effect on
the surface (see Fig. 2) associated with Kondo
breakdown\cite{Alexandrov_PRL2015}. This effect causes the local
moments to decouple from the topological surface states over a wide
temperature range, leading to lighter quasiparticles with large Fermi
surface orbits that resemble the extremal orbits of an unhybridized
bulk, in accordance with ARPES
experiments\cite{Jiang_NatComm2013,Neupane_NatComm2013, Xu_PRB2013,
Frantzeskakis_PRB2013}. In our proposal, the low temperature upturn in
the quantum oscillation amplitudes are most naturally accounted for as
an onset of surface, rather than bulk quantum criticality.


SmB$_6$ provides a unique platform to study the orbital effects of
magnetic fields in narrow gap insulators: it
has a transport gap of about $T_K=$50K and a direct gap of about
230K(20meV).  Other Kondo insulators  (e.g YbB$_{12}$ 
\cite{Okamura_PRB1998} or Ce$_3$Bi$_4$Pt$_3$ ($T_K\sim$35K)\cite{Hundley_PRB1990} are ``Pauli
limited'', with a gap that closes linearly in a field, closing
at a critical field in the range 
$B_{c}\sim 20-50T$.
By contrast, in SmB$_{6}$, the much 
smaller size of the f-electron g-factor ($g\sim \ 0.1-0.2$) 
suppresses the Zeeman splitting in favor of an orbital closure of the
gap at a critical field in excess of $100T$\cite{field2,highfield1}. 


\noindent {\it Bulk quantum oscillations:} We begin by revisiting the
intriguing suggestion that a narrow gap insulator may support bulk quantum
oscillations.  De Haas van Alphen oscillations are a 
thermodynamic effect, resulting from the Landau quantization of a
Fermi surface.  The Lifschitz-Kosevitch temperature dependence of dHvA
oscillations is a result of the discretized sampling of the density of
states.

One interesting possibility is that the observed quantum oscillations
are a magnetic breakdown effect resulting from 
the small hybridization gap. Magnetic breakdown\cite{Cohen_PRL1961, 
Blunt_PR1962, Shoenberg_book} is a result of the breakdown of the quasi-classical 
approximation, where the electrons orbiting around the Fermi surface 
can tunnel among different Fermi surfaces giving rise to larger orbits. 
Such processes preserve energy but not crystal momentum. However, in the 
case of small gap systems, magnetic breakdown leads to tunneling 
through the gap, as a magnetic analog of the Zener breakdown\cite{Kane_JPhysChemSol1959}.
Such effects are known to occur in
type-II superconductors and\
Knolle and as
Cooper\cite{Knolle_Arxiv2015} point out, may be also be a feature 
of narrow gap Kondo insulators, independently of whether they
are topological. It would be interesting to test this
idea in other small gap Kondo insulators like YbB$_{12}$ and
Ce$_4$Bi$_3$Pt$_4$.
Here, the basic idea is that if the cyclotron energy $\hbar \omega_{c}$ is at least
comparable with the direct hybridization gap $V$, $\hbar
\omega_{c}\gtappr V$ Landau quantization
develops, producing a kind of quantum Hall insulator. 
The
authors also  
point out that the oscillation amplitude in this case can deviate from 
standard Lifshitz-Kosevich formula
if the chemical potential is close to the valence or conduction bands.
On the other hand, if cyclotron frequency is much smaller 
than the direct gap, $\hbar \omega_c \ll V$, the oscillation amplitude is
damped exponentially, $R\sim R_0 \exp(-V/\hbar \omega_c)$.

However,  the field range used to observe quantum oscillations
SmB$_6$ lies in the small gap limit $\hbar
\omega_{c}<<V$: to see this, note that at $B=10$T, the
cyclotron frequency is $\hbar \omega_c \simeq 1-2$ meV\footnote{$\hbar \omega_c(B=10T) \simeq 
1-2$ meV is estimated from fitting the tight binding model to a parabolic band. We have taken the band width of the tight binding model 
to be 4 eV in accordance with electronic structure calculations\cite{Fu_PRL2013}. Moreover for a LaB$_6$\cite{Arko_PRB1976}, which has a similar conduction band without the local moments, the effective smallest effective mass $m^*=0.2m_0$ which gives $\hbar \omega_c(B=10T) \simeq 0.5$ meV.}, while
the direct gap measured by
ARPES\cite{Jiang_NatComm2013,Neupane_NatComm2013, Xu_PRB2013,
Frantzeskakis_PRB2013} and optics\cite{Gorshunov_PRB1999} is an order
of magnitude larger, around $V= 20$meV, which would lead to an
exponential suppression of the
amplitude relative to the Lifshitz-Kosevich formula. 
There is a well-known discrepancy between the direct gap
measured in ARPES and optics and the much smaller transport gap 
of about 50K ($\sim 4.3$meV) measured from the resistivity, which may
be associated with an indirect gap. However, it is the direct gap  
$V$ that controls the presence of Landau quantization, and so
long as $\hbar  \omega_{c}\ll V$,
magnetic breakdown is unable to account for the observation of
Lifschitz Kosevitch behavior over a wide temperature and field range in
SmB$_{6}$. 

Another intriguing idea, is that a bulk insulator
might contain a Fermi surface of non-conducting excitations 
that can still develop Landau quantization. This is the essence
of a recent proposal by Baskaran\cite{Baskaran_Arxiv2015}.
Baskaran's work raises the 
fascinating question of whether a Fermi surface of electrically neutral particles can still
experience a Lorentz force. 
Quasiparticles in Landau orbits circulate in quasi-classical orbits,
which for SmB$_{6}$ can be 
as large as 1 micron\cite{Sebastian_Science2015}, which sets a minimum
size for the regions of gapless excitations in the bulk, or on the
surface.  Landau quasiparticles lead to diamagnetism and hence carry 
circulating currents, which in turn implies the quasiparticles must couple
to the current operator ${\bf j}$. The
problem is that a diamagnetic coupling inevitably implies a coupling to the 
electric field ${\bf E}$. 

To demonstrate this point, consider 
a thought experiment in which the magnetic field $\bf{B}$ on 
Landau-quantized quasiparticles is adiabatically
increased, raising the energy of each Landau level as required to
produce quantum oscillations. 
The energy of the single particle states in the n-th level increases 
by an amount of order $n\hbar \Delta \omega_{c}$ as a result of the
change in the magnetic field. Microscopically, the coupling of the
Hamiltonian occurs via the vector potential, 
so the change in ehergy is
given by:
\begin{eqnarray}
n\hbar \Delta 
\omega_c &=&\int \langle \psi_{H}^{(n)} | \frac {\partial H}{\partial t} | \psi_H^{(n)} \rangle dt   \nonumber \\
&=&\int
\langle \psi_{H}^{(n)} | -\frac {\delta  H}{\delta {\bf A} (x)}
| \psi_H^{(n)} \rangle\cdot  \left( 
-\frac{\partial {\bf A} (x)}{\partial t}\right)  d^{3}xdt  \nonumber 
\label{eq:1}
\end{eqnarray}
where $|\psi_H^{(n)}\rangle$ is the one-particle 
wave function of a quasiparticle in the  $n^{th}$ 
Landau level, written in the Heisenberg representation.
But the derivative of the vector potential can be identified with the
electric field ${\bf E}= -\frac{\partial{\bf A}}{\partial t}$, while
$-\frac {\delta  H}{\delta {\bf A} (x)}= {\bf j} (x) $ is the current
operator, so that 
\begin{equation}\label{}
n \hbar \Delta \omega_c = \int \langle \psi_{H}^{(n)} | {\bf j } (x) | \psi_H^{(n)} \rangle\cdot {\bf E} (x)~  d^{3}xdt 
\end{equation}
From this argument, we
see that the energy deposited into quasiparticles 
due to ramping up the field derives from the electric 
field associated with a changing vector potential, i.e Faraday's law.  
Moreover, in order that the field increase the
particles energy, the current operator of the quasiparticles must be
finite. Gauge invariance tells us that since 
${\bf E}= - \nabla \phi - \frac{\partial {{\bf A}}}{\partial t}$, 
the quasiparticles must couple equally to 
an electric field induced by Faraday's effect or a gradient of the
potential, i.e they must be charged and will be conducting. 
Baskaran proposes\cite{Baskaran_Arxiv2015} that a non-conducting Fermi
surface of Majorana fermions can Landau quantize. However, propagating 
Majorana fermions carry a equal weight of electrons and holes moving
in the same direction, so their current matrix elements vanish and
they are fully electrically neutral. 
Thus, as far as we can see, a single band of Majorana fermions
can not Landau quantize and will not give rise to Quantum
oscillations. 

\noindent {\it Surface quantum oscillations:} 
The difficulties encountered in a bulk interpretation of the 
quantum oscillations in insulating, SmB$_{6}$ thus lead us to
re-examine the possibility that these signals are a topological
surface effect. 
At first sight, such a proposal should be ruled out because
the Fermi surface area of the topological surface states are unrelated to
the bulk Fermi surface of the conduction electrons in simple model
theories for topological insulators. Indeed, the surface Fermi
surface area can be vanishingly small if the chemical potential is
close to (or at) the Dirac point. Nonetheless, SmB$_6$ offers us some
empirical support for a surface interpretation of the oscillations, 
for ARPES measurements\cite{Jiang_NatComm2013,Neupane_NatComm2013,
Xu_PRB2013, Frantzeskakis_PRB2013} show that the Dirac point is sunk
into the valance band and regardless of where the surface chemical
potential is, the Fermi surface area of the topological surface states
are quite close to bulk Fermi surface area\cite{Legner_arXiv2015}. 
Moreoever, as shown 
Fig. 1, the low frequency oscillations of both experiments agree
within the errorbars, indicating that, at least for these modes, the 
2D surface states interpretation deserves further consideration. 

The empirical 
observation that the  Dirac point of the surface states
is sunk in the valence band suggests a strong
particle-hole asymmetry. One mechanism for this asymmetry is through
the destruction of Kondo singlets at the
surface\cite{Alexandrov_PRL2015}. 
The different parity of the localized f-electrons
and mobile d-electrons in SmB$_6$, 
causes Kondo screening to develop by coupling to 
nearest neighbor sites. 
The reduction of nearest neighbor
sites on the surface causes the surface Kondo temperature to be 
suppressed, so that the screening of local moments at the surface is
either develops at much lower temperatures, or fails completely due
to the intervention of surface magnetic order or quantum criticality. 
Such {\sl Kondo breakdown}
liberates conduction electrons which were bound in Kondo singlets,
which dopes the topological surface states and pushes the Dirac point
to the valence band. The resulting surface states have a considerably 
larger Fermi
surface given by 
\begin{eqnarray}
\frac{\mathcal{A}_{\rm FS}}{(2\pi)^2} = \Delta n_f	
\end{eqnarray}
in accordance with Luttinger's sum rule where $\Delta n_f$ is the
change of the Samarium nominal valence on the surface due to the
re-localization of the f-electrons. Note that the nominal valence, a 
concept introduced by Anderson\cite{Anderson_book}, is different 
than real valence measured by photoemission. In the Kondo insulating 
state, the nominal valence of Sm is $+2$ whereas the real valence is
$+2.6$. Kondo breakdown also leads to much lighter
quasiparticles (see Fig 2(d)), giving rise to a Fermi velocity and a
large 2D Fermi surface area which are in agreement with ARPES
experiments\cite{Jiang_NatComm2013,Neupane_NatComm2013, Xu_PRB2013,
Frantzeskakis_PRB2013}. Indeed, Kondo breakdown gives rise
to topological surface states which mirror the extremal orbits 
of the bulk d-states. 
This leads us to propose that the 
quantum oscillations observed in both experiments are a consequence of
Kondo breakdown extending down to involve 
to about 1K, following the Lifshitz-Kosevich formula for a large,
light Fermi surface.

Next we consider the case when a Kondo effect starts to develop at
the surface. The low energy Hamiltonian of the surface moments 
interacting with topological surface states is given by a chiral Kondo 
or Anderson model\cite{Alexandrov_PRL2015}. We use the large $N$ description of chiral Anderson model:
\begin{eqnarray}
H_{CAM} &=& \sum_{k} (v_F{\bf k} \cdot \sigma_{\alpha \beta}-\mu) c_{\bk \alpha}^\dagger c_{\bk \beta}  \nonumber \\
&&+ \sum_k (\epsilon_{f}({\bk} ) b_s^2+\lambda) f_{\bk \alpha}^\dagger f_{\bk \alpha}  \nonumber \\
&&+ \sum_k (V_z \sigma^z_{\alpha \beta} + V_{\perp} {\bf k} \cdot \sigma_{\alpha \beta}) b_s f^\dagger_{k\alpha} c_{\bk\beta} \nonumber \\
&&+\lambda \sum_i(1+b_s^2-n_f)
\end{eqnarray}
where $c$ and $f$ are fermonic operators for the topological
surface states and f-electrons. The momentum, ${\bf k}$ is defined in the 
two dimensional surface Brillouin zone, here it is in $xy$ plane. $v_F$ and $b_s$ are the Fermi velocity of
the conduction electrons and the slave-boson surface
amplitude. $\epsilon_{f}({\bk})$ and $\lambda$ are the dispersion and the
effective chemical potential of the $f$ band. The explicit form of $\epsilon_{f}({\bk})$
is not important yet nearest neighbor hopping gives 
$\epsilon_{f}({\bk})=2t_f (\cos k_x+\cos k_y)$ where $t_f$ is the bare $f$ hopping amplitude. 
$\epsilon_{f}({\bk})$, which is usually ignored in simple Kondo models is crucial in this case 
to open up an indirect gap.
Since the inversion symmetry is broken along $z$ direction, previously forbidden on-site
hybridization is now allowed. This model provides a minimal
description of the surface Kondo effect between the spin-orbital
polarized surface states and doubly degenerate surface f-states. In
its limiting forms, the 
ground-state of the chiral Anderson model will be either magnetically
ordered state  or a heavy fermion liquid, but intermediate states
involving quantum criticality, fractinalized 
spin-liquids\cite{thomsonsachdev} and 
superconductivity might form. However, at this stage our simple mean-field
theory merely predicts a reduced surface Kondo temperature.

Indications of a revival of surface Kondo effect at low temperatures
have been observed around few Kelvin in magnetothermoelectric
transport experiments\cite{Luo_PRB2015} where the surface states
become much heavier. As shown in Fig. 2 (a-d), theory predicts that as
the Kondo effect develops at the surface, the doubly degenerate
$f$-band hybridizes with the light, spin-orbital polarized surface
states. As a result, only one component of the f-band can hybridize
and the other, unhbyridized band cuts remains gapless, forming a new
Dirac point at the high symmetry point. During this process part of
the Fermi surface turns hole-like from electron-like (Fig. 2 (a-d)
insets), to account for the change of the total carrier
density. Nevertheless, the large Fermi surface orbit is preserved up
to large values of $b_s^2$. The moment Kondo effect turns on $b_s
\rightarrow 0_+$, the effective mass diverges and becomes negative for
the large Fermi surface for $b_s^2\neq 0$. The observation of these
new multiple Fermi surface orbits at intermediate temperatures is an
important test of the theory.

\noindent {\it Discussion:} One likely explanation for 
the upward deviation in the temperature dependence of the quantum
oscillations from the
Lifshitz-Kosevich formula is the development of {\sl surface}
quantum criticality. 
On both experimental\cite{Lohneysen1996,
Julian1996} and theoretical grounds\cite{Pixley_Arxiv2015}, the
suppression of the Kondo effect by 
magnetism is known to dramatically enhance  quasiparticle masses.
Moreover \cite{Sebastian_Science2015} reports that
the oscillation amplitudes fits well with theories of quantum critical
metals. Another possibility for the deviation from the
Lifshitz-Kosevich formula might be 
surface magnetic break-down. The hybridization 
gap at the surface is significantly 
smaller than the bulk gap $V_s \sim 0.1-1$ meV. In this situation, the
Knolle-Cooper\cite{Knolle_Arxiv2015} magnetic break-down 
mechanism is expected to become active,  
since $\hbar \omega_c > V_s$. In this situation, the Landau quantized
orbits are not affected by the small gap and are expected to give rise
to quantum oscillations with a deviation from the Lifshitz-Kosevich
formula\cite{Knolle_Arxiv2015}. 

One of the issues that our discussion is unable to address, is 
the long-standing issue of a linear temperature dependence in the
specific heat, where a conservative estimate
of the linear specific heat $C_{V}/T=\gamma =$ 2mJ/mol/K$^{2}$ is at
least 
twice that of bulk metallic LaB$_{6}$. 
One interpretation of this specific heat 
might ascribe it to a neutral Fermi surface, even though as we have
discussed,  such quasiparticles can not exhibit Landau quantization. 
Although past thermal transport
experiments\cite{Flachbart_1982} reported a $T^3$ rather than a
$T$-linear thermal conductivity, the recent developments 
surely warrant repeating these measurements on 
higher quality samples.  Another possibility for the linear specific
heat is inhomogenous metallic bubbles which could be probed by
neutrons and muon-spin resonance experiments. 

\noindent
{\it Conclusion:} We have contrasted the 
SmB$_6$ quantum oscillation data that has inspired different interpretations.
After discussing the proposed three-dimensional, bulk 
interpretations of the oscillations, we have been led to propose an
alternative topological surface interpretation, modified by the
effects of Kondo breakdown. 
In this picture the deviation from
Lifshitz-Kosevich formula at low temperatures may either be a result
of surface quantum criticality, or the suppressed low-temperature
onset of the surface Kondo effect. 
Since the temperature dependence of the high frequency
data is qualitatively the same as the low
frequency\cite{Sebastian_PriComm}, it is tempting to also ascribe 
the high frequency oscillations to 
topological surface states, most likely the 2D Fermi surfaces 
located around the X point. A more careful angle
dependence of the oscillation amplitude is required to test this
hypothesis. 
Lastly, we note that if the surface Kondo effect does indeed set in at
low temperatures, a set of new, high mass
frequencies should develop at low
fields and low temperatures. 

\noindent

{\it Acknowledgments:} We gratefully acknowledge stimulating
conversations with Jim Allen, Luis Balicas, Gilbert Lonzarich, Lu Li,
Eugene Mele and Suchitra Sebastian. This work is supported by Department of Energy grant
DE-FG02-99ER45790.

{\it Note added:} We have recently became aware of another calculation\cite{Zhang} that uses
magnetic breakdown mechanism similar to Knolle and Cooper\cite{Knolle_Arxiv2015}.

\bibliographystyle{apsrev}
\end{document}